\documentclass[12pt,preprint]{aastex}

\begin{document} 
\title{CORE MASS FUNCTION: THE ROLE OF GRAVITY}

\author{Sami Dib\altaffilmark{1,2,4}, Axel Brandenburg\altaffilmark{3}, Jongsoo Kim\altaffilmark{4}, Maheswar Gopinathan\altaffilmark{4,5}, and Philippe Andr\'{e}\altaffilmark{1}}

\altaffiltext {1} {Service d'Astrophysique, DSM/Irfu, CEA/Saclay, F-91191, Gif-sur-Yvette Cedex, France; sami.dib@cea.fr, pandre@cea.fr}
\altaffiltext {2} {Lebanese University, Faculty of Sciences, Department of Physics, El Hadath, Beirut, Lebanon}
\altaffiltext {3} {NORDITA, Roslagstullsbacken 23, AlbaNova University Center, 10691 Stockholm, Sweden; brandenb@nordita.org}
\altaffiltext {4} {Korea Astronomy and Space Science Institute, 61-1, Hwaam-dong, Yuseong-gu, Daejeon 305-348, Korea; jskim@kasi.re.kr, maheswar@kasi.re.kr}
\altaffiltext {5} {Aryabhatta Research Institute of Observational Sciences, Manora Peak, Nainital 263129, India}

\begin{abstract} 
We analyze the mass distribution of cores formed in an isothermal, magnetized, turbulent, and self-gravitating nearly critical molecular cloud model. Cores are identified at two density threshold levels. Our main results are that the presence of self-gravity modifies the slopes of the core mass function (CMF) at the high mass end. At low thresholds, the slope is shallower than the one predicted by pure turbulent fragmentation. The shallowness of the slope is due to the effects of core coalescence and gas accretion. Most importantly, the slope of the CMF at the high mass end steepens when cores are selected at higher density thresholds, or alternatively, if the CMF is fitted with a lognormal function, the width of the log-normal distribution decreases with increasing threshold. This is due to the fact that gravity plays a more important role in denser structures selected at higher density threshold and leads to the conclusion that the role of gravity is essential in generating a CMF that bears more resemblance with the IMF when cores are selected with an increasing density threshold in the observations.      

\end{abstract} 


\section{INTRODUCTION}\label{intro}
It is now widely accepted that stars form in dense and gravitationally bound molecular cloud (MC) cores. Thus, understanding the properties of the core mass function (CMF) is the cornerstone of any successful theory of the stellar initial  mass function (IMF). The question whether there is a 1-to-1 mapping between the CMF and the IMF remains an open one. The CMF is usually represented by $dN/dM=M^{\alpha}$, where $\alpha$ takes different values in different mass ranges (for the IMF, $\alpha=-2.35$ for $M \gtrsim 0.5$ M$_{\odot}$, Salpeter 1955). The determination of the CMF of cores of several MC regions have been obtained using a variety of wavelengths and techniques such as mm spectroscopy of various molecular lines, submm/mm continuum emission from dust, and stellar near infrared absorption by dust. Usually, no prior assumptions are made on the type of structures identified in these studies. They can be a mixture of prestellar cores, unbound objects, and eventually protostellar envelopes. For nearby MCs, where it is possible to identify dense cores (with average number densities $n \gtrsim 10^{4}$ cm$^{-3}$; hereafter Cores), the CMF is usually described by a two component power law above and below a turnover point (with exponents $\alpha_{h}$ and $\alpha_{l}$, respectively). Starting with the 1.2 mm study of the Ophiuchus starless dense cores by Motte et al. (1998), a number of submm/mm dust continuum observations have found $\alpha_{h}$ and $\alpha_{l}$ to be in the range [-2.1,-3] and [-1.3,-1.5], respectively (Motte et 2001; Testi \& Sargent 2001; Johnstone et al. 2006; Beuther \& Schilke 2004; Stanke et al. 2006; Reid \& Wilson 2006; Johnstone \& Bally 2006; Kirk et al. 2006; Enoch et al. 2006). These estimates bracket the values of the exponents of the IMF (Scalo 1998; Kroupa 2002; Chabrier 2003). Similar slopes are found using high density molecular tracers such as H$^{13}$CO$^{+}$ (Onishi et al. 2002; Ikeda et al. 2007) and the near infrared absorption technique (Alves et al. 2007). On the other hand, submm/mm observations of more distant clouds in which it is only possible to identify lower density structures ($n \sim 10^{2}-10^{3}$ cm$^{-3}$; hereafter Clumps), exhibit values of $\alpha_{h}$ in the range of [-1.4,-1.8] if the whole CMF is fitted with a single power law (Coppin et al. 2000; Kerton et al. 2001; Tothill et al. 2002; Mookerjea et al. 2004; Massi et al. 2007; Li et al. 2007, Mu\~{n}oz et al. 2007) or slightly steeper ($\alpha_{h} \sim -2$) if it can be fitted by a two component power law (Reid \& Wilson 2005). Similar slopes are obtained from molecular line observations using low/intermediate density molecular line tracers (Stutzki \& G\"{u}tsen 1990; Lada et al. 1991; Blitz 1993; Brand \& Wouterloot 1995; Kramer et al. 1998; Heithausen et al. 1998; Kawamura et al. 1998; Wilson et al. 2003). Values of $\alpha_{h} \sim [-1.6,-1.7]$ resemble the slopes of the stellar clusters mass function (Elmegreen 2000), which hints to the fact that these objects cannot be the direct progenitors of individual stars and must undergo additional fragmentation.  
 
The discrepancy in $\alpha_{h}$ between the Cores and Clumps populations does not seem to be merely an effect of the spatial resolution and hence of mass resolution. Additional evidence might be that, in Motte et al. (1998), the inclusion of the larger size cores in the CMF changes the values of $\alpha_{h}$ from $-2.5$ to $\sim -1.5$. An explanation of the origin of the discrepancy would be that different physical processes dominate the evolution of these two populations of cores. The close resemblance between the Cores CMF and the IMF suggests that they are more likely to be gravitationally bound (which is confirmed by mm line observations, Andr\'{e} et al. 2007) and are the direct progenitors of stars or small stellar systems, whereas Clumps are more likely to be dominated by turbulence. Note that the description of the CMF by power law functions is not the unique option. Reid el al. (2006) argued that a lognormal function is a better fit to the observed CMF in many regions and Goodwin et al. (2007) showed that the CMF of the submm cores obtained by Nutter \& Ward-Thompson (2007) is well represented by a lognormal distribution (also Fig. 2 in Andr\'{e} et al. 2008). 

 Several physical processes are believed to play a role in shaping the CMF: gravitational and thermal fragmentation (Larson 1985; Klessen 2001; Li et al. 2003), supersonic turbulence (Padoan \& Nordlund 2002, PN02), accretion (Larson 1992; Bonnell et al. 2001; Bate \& Bonnell 2005), core coalescence (Murray \& Lin 1996; Dib et al. 2007a,2008), feedback (Adams \& Fatuzzo 1996), and magnetic fields (Shu et al. 2004). A number of numerical studies of MCs that include some of these physical processes have derived a CMF that bears some resemblance with the CMF of Cores and the IMF (Gammie et al. 2003; Li et al. 2004; Ballesteros-Paredes et al. 2006; Padoan et al. 2007) or with the CMF of Clumps (Hennebelle \& Audit 2007, Heitsch et al. 2008). The origin of the differences between the Cores and Clumps CMFs and the eventual link between them has not been investigated so far. In this study, we identify cores in a high resolution 3D numerical simulation of a magnetized, turbulent, and self-gravitating MC model at different density thresholds. In Dib et al. (2007b), we have shown, using a detailed virial analysis that the inner parts of dense structures, identified at higher density thresholds, are more gravitationally bound than the same objects identified at lower thresholds. Thus, constructing the CMF at different thresholds is a direct test of the effects of gravity on the CMF. A striking example of how the density threshold can modify the inferred masses of cores is the case of the core L1512. Its derived mass from CO observations is $\sim 6$ M$_{\odot}$ (Falgarone et al. 2001), whereas it was estimated from submm observations at $\sim 1.4$ M$_{\odot}$ (Ward-Thompson et al. 1999). For the whole CMF, Kramer et al. (1998) derived a value of $\alpha_{h}=-1.72 \pm 0.09$ for the Orion B-S region from $^{13}$CO(1-0) observations whereas Johnstone et al. (2006) derived a value of $\sim -2.5$ from submm observations. Another example is the case of the NGC 7538 region, observed by Kramer et al. (1998) to have a CMF slope of $\alpha_{h}=-1.65 \pm 0.05$ in the $^{13}$CO(1-0) line and $\alpha_{h}=-1.79 \pm 0.12$ in the C$^{18}$O(1-0) line which traces higher density material. 

\section{THE SIMULATIONS, CLUMP-FINDING ALGORITHM, AND ANALYSIS STRATEGY}\label{simul}
We performed a 3D numerical simulation of a magnetized, self-gravitating, and turbulent isothermal MC using the TVD code (Kim et al. 1999). The simulation is similar to the ones presented in Dib et al. (2007b) but with a higher numerical resolution of $512^{3}$ grid cells. In Dib et al. (2007b), we found that the population of cores formed in a moderately magnetically supercritical MC reproduces best some of the available observations such as the virial parameter-mass relation. Thus, we here only consider this cloud model for further analysis. The features of this simulation are: Turbulence is driven until it is fully developed (for 2 crossing timescales) before gravity is turned on. The Poisson equation is solved to account for the self-gravity of the gas using a standard Fourier algorithm. Turbulence is constantly driven in the simulation box following the algorithm of Stone et al. (1998). The kinetic energy input rate is adjusted such as to maintain a constant rms sonic Mach number $M_{s}=10$. Kinetic energy is injected at large scales, in the wave number range $k=1-2$. Periodic boundary conditions are used in the three directions. In physical units, the simulation has a linear size of 4 pc, an average number density of $\bar{n}=500$ cm$^{-3}$, a temperature of 11.4 K, a sound speed of $0.2$ km s$^{-1}$, and an initial {\it rms} velocity of 2 km s$^{-1}$. The Jeans number of the box is $J_{box}=4$ (number of Jeans masses in the box is $M_{box}/M_{J,box}=J_{box}^{3}=64$, where $M_{box}=1887$ M$_{\odot}$). The initial magnetic field strength in the box is $B_{0}= 14.5$ $\mu$G. Correspondingly, the plasma beta and mass-to-magnetic flux (normalized for the critical value for collapse $M/\phi =(4 \pi^{2} G)^{-1/2}$, Nakano \& Nakamura 1978) values of the box are $\beta_{p,box}=0.1$, and $\mu_{box}=2.8$. Cores are identified using a clump-finding algorithm that is based on a density threshold criterion and a friend-of-friend approach as described in Dib et al. (2007b). We restrict our selection of cores to epochs where the Truelove criterion (Truelove et al. 1997) is not violated in any of them. Thus, our derived CMF applies to starless prestellar cores. At a given threshold, the number of cores detected at one epoch is insufficient to sample appropriately the whole mass spectrum. It is therefore necessary to combine cores selected at several epochs in order to construct a CMF that is statistically significant. We combine the populations of cores identified at 35 timesteps. This approach resembles the situation in the observations where cores identified in different locations of a MC are most likely in different stages of their evolution. Even for a composite multi-epoch spectrum, the statistics of cores identified with thresholds $\gtrsim 60~\bar{n}$ remains poor. 

\section{CORE MASS FUNCTION}\label{spectra}
\begin{figure}
\plotone{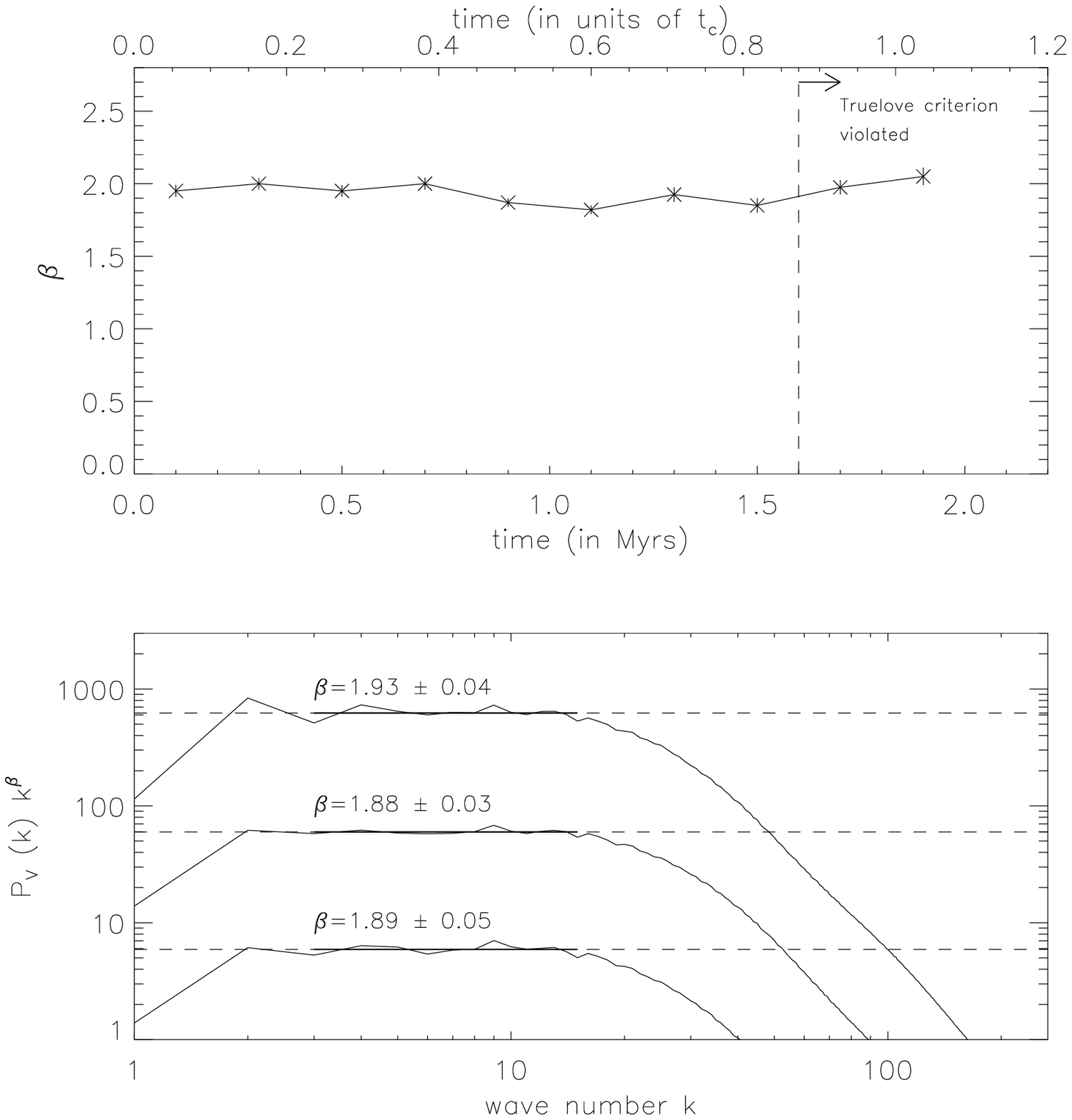}
\caption{Time evolution of the exponent of the turbulent velocity field power spectrum, $\beta$ (top) after gravity is turned on, and a sample of velocity power spectra at selected epochs (bottom). Each data point in the top figure is an average over five neighboring timesteps.}
\label{fig1}
\end{figure}

\begin{figure}
\plotone{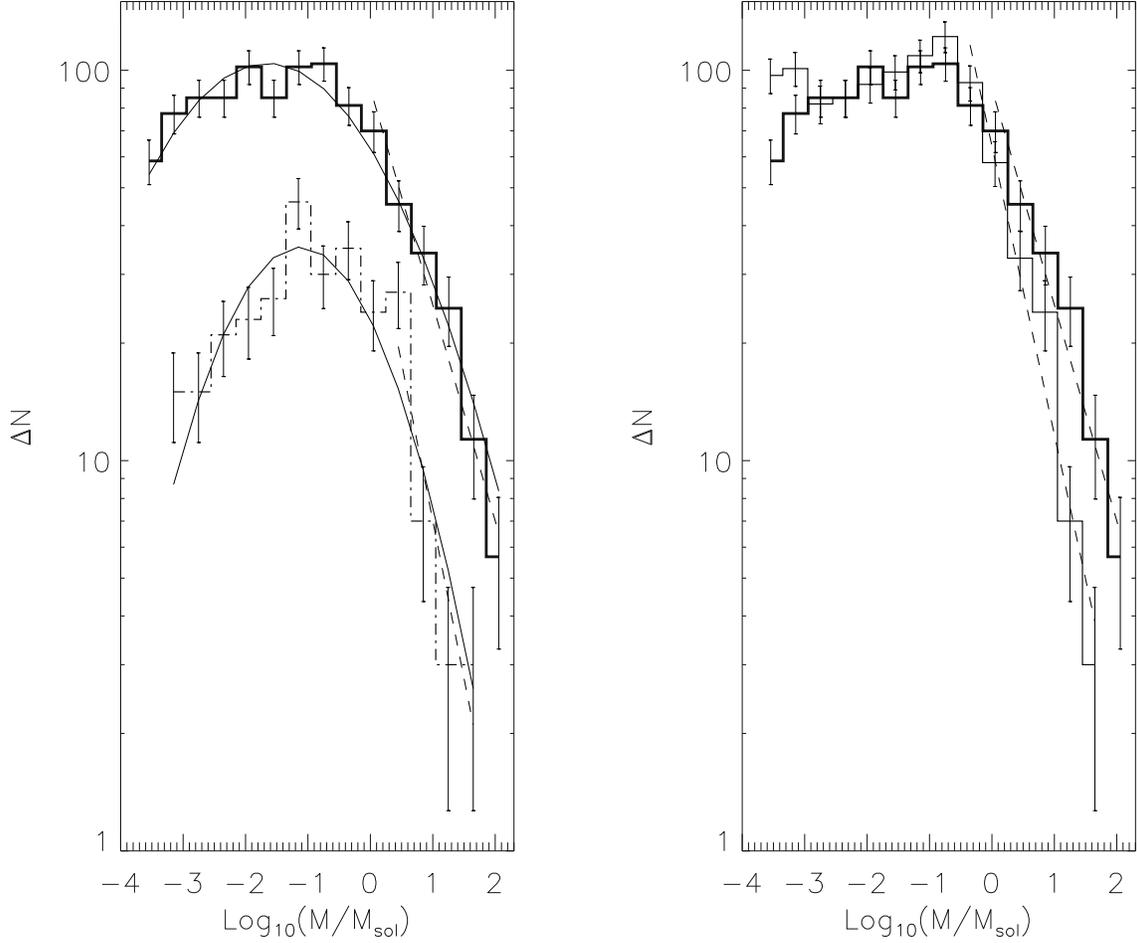}
\caption{Left: The CMF of cores identified in the simulation with gravity at the thresholds of $20~\bar{n}$ (thick full line) and $50~\bar{n}$ (thin dash-dotted line). Overplotted are lognormal fits (thin line) and power law fits to the high mass end (dashed line). Right: The CMF of cores at the threshold of $20~\bar{n}$ for the models with gravity (thick full line) and without gravity (thin full line). Power law fits to the high-mass end are overplotted (dahsed lines).}
\label{fig2}
\end{figure}

Before discussing the CMF obtained from the simulation, it is worth checking what CMF is anticipated from the turbulent fragmentation formalism of PN02. Note that this formalism does not explain the CMF of Clumps. In the PN02 approach, $\alpha_{h}$ is given by $\alpha_{h}= [-3/(4-\beta)]-1$, where $\beta$ is the exponent of the turbulent velocity field power spectrum ($P_{v}(k)=k^{-\beta}$). In order to reproduce the CMF of Clumps, the formalism of PN02 requires $\beta$ to be close to zero, which is not observed. Fig.~\ref{fig1} displays the time evolution of $\beta$ in our simulation, and a few velocity spectra at selected epochs. The value of $\beta$ is observed to fluctuate around $\sim 1.9$. According to PN02, this would result in a value of $\alpha_{h} \sim -2.42$ (at a given epoch and for a multi-epoch built CMF). Fig.~\ref{fig2} (left) displays the CMF for cores identified at the density thresholds of 20 and $50~\bar{n}$. Both samples use the same number of timesteps. Several features of the CMFs can be noted. If the high-mass end of the spectra is to be assimilated to a power law, fits to the CMF in this regime yield a steeper slope for the high density threshold sample than for the one with the lower threshold. The slopes for the $20~\bar{n}$ and $50~\bar{n}$ CMFs are $-1.55 \pm 0.15$ and $-1.80 \pm 0.2$ in the mass ranges [1,113] M$_{\odot}$ and [2.8,44] M$_{\odot}$ respectively. However, the CMFs are also well fitted over their entire mass range by a lognormal distribution whose width is smaller at higher threshold. The width of the lognormal fit of the CMF is $1.63 \pm 0.07$ and $1.12 \pm 0.09$ at the $20$ and $50~\bar{n}$ levels, respectively (the width of the IMF is $0.57$, Chabrier 2003). A two-sided Kolmogorov-Smirnov (KS) test indicates that the $20~\bar{n}$ and $50~\bar{n}$ CMFs have a probability of only 2 percent of being drawn from the same parent distribution, a result significant at the 2.3 sigma level.  For reference, we also calculated the CMF slope of $20~\bar{n}$ cores in a  non self-gravitating model (Fig.~\ref{fig2}, right). The slope of the CMF in the high mass range is larger for the non self-gravitating cloud model than for the self-gravitating one ($\alpha_{h}=-1.73 \pm 0.07$ and resembles the one predicted and obtained by Hennebelle \& Audit (2007) for turbulence dominated, non self-gravitating clumps (in this non self-gravitating case, statistics are not enough to build the $50~\bar{n}$ CMF).
 
\begin{figure}
\plotone{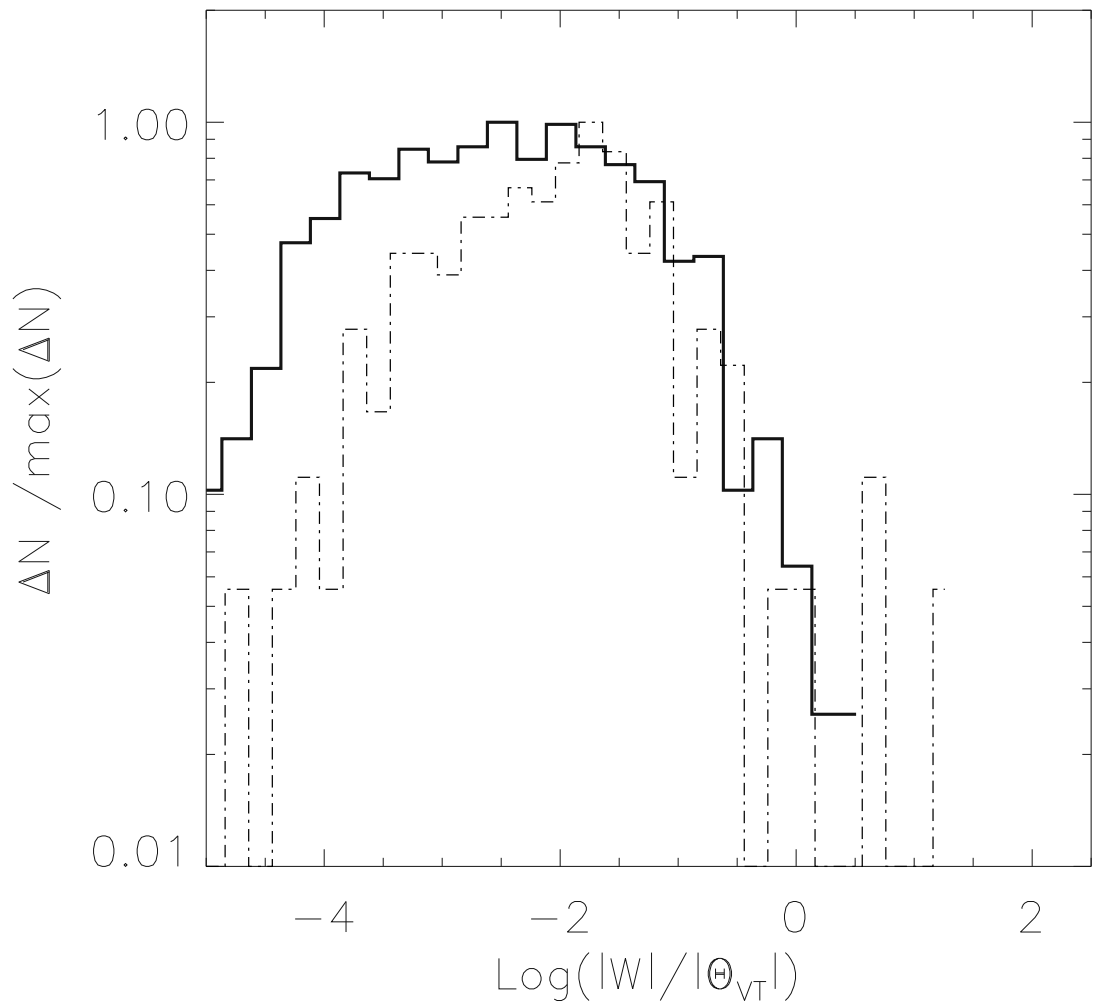}
\caption{Normalized probability distribution function of the ratio of the gravitational term $W$ to both volume and surface energy terms that appear in the virial equation (kinetic, thermal, and magnetic; full line is for the $20~\bar{n}$ cores and the dashed line is for the $50~\bar{n}$ cores). The latter terms are condensed in the quantity $\Theta_{VT}$ (see text for details). The larger $|W|/|\Theta_{VT}|$ is, the stronger the influence of gravity. A KS test of the (non-normalized) distributions shows that they have a probability of 3 percent of being identical.}
\label{fig3}
\end{figure}

\begin{figure}
\plotone{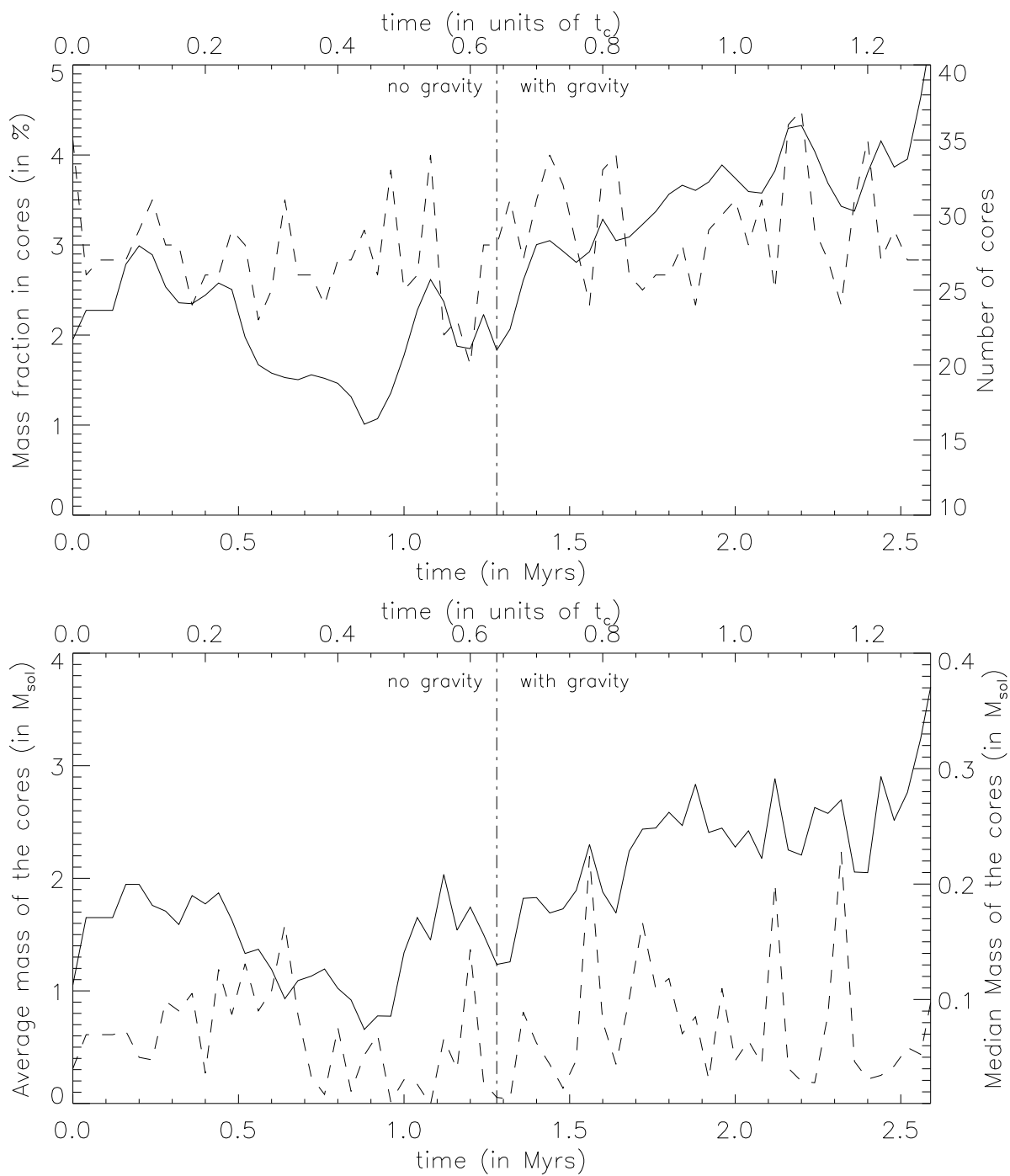}
\caption{Time evolution of: The fraction of the mass present in the cores (top, full line), number of cores (top, dashed line), average mass of the cores (bottom, full line), and median mass of the cores (bottom, dashed line). Time is shown in Myrs (bottom horizontal axis) and in units of the crossing time $t_{c}=2$ Myr (top horizontal axis).}
\label{fig4}
\end{figure}

The slopes of the CMF at high mass end are shallower than those predicted by PN02 and steepen with increasing threshold. We interpret the shallower than the IMF slopes and the steepening with increasing threshold to the role played by gravity in modifying the CMF generated by turbulence. The steepening of the slope with increasing threshold is due to the increasing importance of gravity at higher thresholds. Fig.~\ref{fig3} displays the ratio of the gravitational term in the virial theorem ($W=-\int_{V} \rho r_{i} (\partial \phi/\partial r_{i}) dV$, where $\phi$ is the gravitational potential) to the other terms that appear in the virial equation. The latter terms are condensed in the quantity $\Theta_{VT}=2~(E_{th}+E_{k}-\tau_{k}-\tau_{th})+E_{mag}+\tau_{mag}$, where $E_{th}=\frac{3}{2}\int_{V} P dV$, $E_{k}= \frac{1} {2} \int_{V} \rho_{i} v_{i}^{2} dV$, and $E_{mag}=\frac{1}{8 \pi}\int_{V} B^{2} dV$ are the volume thermal, kinetic, and magnetic energies, respectively, with $P$ being the thermal pressure, and $\tau_{th}=\frac{1}{2}\oint_{S} r_{i} P \hat{n}_{i} dS$, $\tau_{k}=\frac{1}{2} \oint_{S} r_{i} \rho v_{i} v_{j} \hat{n}_{j} dS$, and $\tau_{mag}=\oint_{S} r_{i} T_{ij} \hat{n}_{j} dS$ are the thermal, kinetic, and magnetic surface energies, respectively, where $T_{ij}$ is the Maxwell stress tensor. Fig.~\ref{fig3} indicates that gravity is, on average, more important in the virial balance for the cores defined at the $50~\bar{n}$ threshold which are a mixed population of gravitationally bound and unbound objects than for the $20~\bar{n}$ cores which are essentially unbound objects. An extrapolation to higher thresholds of the order of 500 $\bar{n}=2.5 \times 10^{5}$ cm$^{-3}$ which are number densities similar to those of gravitationally bound cores observed in the submm/mm or in dense molecular tracers would yield slopes in the range [-2.3,-2.5], thus offering a natural explanation for the similarity of the observed slopes between the CMF of dense cores and of the IMF. The reason why the CMF of the $20~\bar{n}$ cores in the self-gravitating regime is shallower than that of the same population in the non self-gravitating case is most likely due to the long range effects of gravity. Fig.~\ref{fig4} displays the time evolution, for the $20~\bar{n}$ cores, of the fraction of the mass present in the cores, number of cores, and average and median masses of the cores. The increasing mass fraction and average mass of the cores at roughly equal number of cores even before any of them proceeds to collapse are a clear indication of mass growth affecting the cores due to the combined effect of gas accretion and core coalescence (see Fig. 16 in Dib et al. 2007b). On the other hand, the constancy of the number of cores and their median mass means that newborn cores are generated with the smallest masses. Although it is not possible to assess from Fig.~\ref{fig4} the contribution of each of these two processes, this mass growth which constitutes a deviation from the purely turbulent regime is responsible for the flattening in the $20~\bar{n}$ CMF in the self-gravitating case as compared to the non self-gravitating one. 

\section{CONCLUSIONS}\label{conc}
In this work, we have constructed the mass function of cores in a 3D numerical simulation of a magnetized, turbulent, and self-gravitating molecular cloud. Cores are identified at the density threshold levels of $10^{4}$ and $2.5 \times 10^{4}$ cm$^{-3}$. In agreement with observations, the slope of the core mass function is found to be steeper for cores identified at the higher threshold level. The steepening of the slope with increasing threshold is due to the increasing importance of gravity in the virial balance of the cores at higher densities and a sign of the role played by gravity in the gravo-turbulent scenario of star formation. On the other hand, at lower thresholds, the slope is shallower than what is found for the purely turbulent case. This is due to the effects of mass growth that affect the cores. This mass growth is essentially due in the initial phase to the effect of core coalescence which generates more massive structures particularly for cores defined at lower thresholds since they have larger cross sections.
     
\acknowledgements
We thank the Referee and Patrick Hennebelle for constructive suggestions. Simulations were run on a high performance cluster at KASI. J.K. is supported  by the Korea Science and Engineering Foundation through the Astrophysical Research Center for the Structure and Evolution of the Cosmos and the grant of the basic research program R01-2007-000-20196-0.
  
{}

\begin{thebibliography}{}
\bibitem[Adams (1996)] {adams96} Adams, F. C., \& Fatuzzo, M. \ 1996, \apj, 464, 256 
\bibitem[Alves (2007)] {alves07} Alves, J., Lombardi, M., \& Lada, C. J. \ 2007, \aap, 462, L17
\bibitem[Andre (2007)] {andre07} Andr\'{e}, P., Belloche, A., Motte, F., \& Peretto, N. \ 2007, \aap, 472, 519  
\bibitem[Andre (2008)] {andre08} Andr\'{e}, P., Basu, S.,  \& Inutsuka, S. in Structure Formation in Astrophysics, Ed. G. Chabrier, (arXiv:0801.4210)
\bibitem[Ballesteros (2006)] {ballesteros06} Ballesteros-Paredes, J., et al. \ 2006, \apj, 637, 384 
\bibitem[Bate (2005)] {bate05} Bate, M. R., \& Bonnell, I. A.  \ 2005, \mnras, 356, 1201 
\bibitem[Beuther (2004)] {beuther04} Beuther, H., \& Schilke, P. \ 2004, Science, 303, 1167 
\bibitem[Blitz (1993)] {blitz93} Blitz, L. \ 1993, in Protostars and Planets III, ed. E. H. Levy \& J. I. Lunine (Tuscon:Univ. Arizona Press), 125
\bibitem[Bonnell (2001)] {bonnell01} Bonnell, I., Clarke, C. J., Bate, M. R., \& Pringle, J. E. \ 2001, \mnras, 324, 573
\bibitem[Brand (1995)] {brand95} Brand, J., Wouterloot, J. G. A. \ 1995, \aap, 303, 851
\bibitem[Chabrier (2003)] {chabrier03} Chabrier, G. \ 2003, \pasp, 115, 763
\bibitem[Coppin (2000)] {coppin00} Coppin, K. E. K., Greaves, J. S., Jenness, T., \& Holland, W. S. \ 2000, \apj, 356, 1031 
\bibitem[Dib (2007a)] {dib07a} Dib, S., Kim, J., Shadmehri, M. \ 2007a, \mnras, 381, L40 
\bibitem[Dib (2007b)] {dib07b} Dib, S., Kim, J., V\'{a}zquez-Semadeni, E., et al. \ 2007b, \apj, 661, 262  
\bibitem[Dib (2008)] {dib08} Dib, S., Shadmehri, M., Maheswar, G., et al. \ 2008, in Massive Star Formation: Observations confront Theory, ASP Conf. Series, (arXiv:0710.3969)
\bibitem[Elmegreen (2000)] {elmegreen00} Elmegreen, B. G., et al.  \  2000, in Protostars and Planets IV, ed. V. Mannings, A. P. Boss, \& S. S. Russel (Tuscon:Univ. Arizona Press), 179  
\bibitem[Enoch (2006)] {enoch06} Enoch, M. L., et al. \ 2006, \apj, 638, 293
\bibitem[Falgarone (2001)] {falgarone01} Falgarone, E., Pety, J., \& Phillips, T. G. \ 2001, \apj, 555, 178 
\bibitem[Gammie (2003)] {gammie03} Gammie, C. F., Lin, Y.-T., Stone, J., M., \& Ostriker, E. C.  \ 2003, \apj, 592, 203
\bibitem[Goodwin (2008)] {goodwin08} Goodwin, S. P., Nutter, D., Kroupa, P., et al.  \ 2008, \aap, 477, 823
\bibitem[Heithausen (1998)] {heithausen98} Heithausen, A., Bensch, F., Stutzki, J., et al.  \ 1998, \aap, 331, L65
\bibitem[Heitsch (2008)]  {heitsch08} Heitsch, F.,  et al.  \ 2008, \apj, 674, 316
\bibitem[Hennebelle (2007)]  {hennebelle07} Hennebelle, P.,  \& Audit, E. \ 2007,  \aap, 465, 431 
\bibitem[Ikeda (2007)] {ikeda07} Ikeda, N., Sunada, K., \& Kitamura, Y. \ 2007, \apj, 665, 1194  
\bibitem[Johnstone (2006a)] {johnstone06a} Johnstone, D., Matthews, H., \& Mitchell, G. F. \ 2006, \apj, 639, 259
\bibitem[Johnstone (2006b)] {johnstone06b} Johnstone, D., \& Bally, J. 2006, \apj, 653, 383
\bibitem[Kawamura (1998)] {kawamura98} Kawamura, A., Onishi, T., Yonekura, et al. \ 1998, \apjs, 117, 387 
\bibitem[Kerton (2001)] {kerton01} Kerton, C. R., Martin, P. G., Johnstone, D., \& Ballantyne, D. R. \ 2001, \apj, 552, 601
\bibitem[Kim (1999)] {kim99} Kim, J., Ryu, D., Jones, T. W., \& Hong, S. S. \ 1999, \apj, 514, 506
\bibitem[Kirk (2006)] {kirk06} Kirk, H., Johnstone, D., \& Di Francesco, J. \ 2006, \apj, 1009, 1023 
\bibitem[Klessen (2001)] {klessen01} Klessen, R. S. \ 2001, \apj, 556, 837
\bibitem[Kramer (1998)] {kramer98} Kramer, C., Stutzki, J., R\"{o}rig, R., \& Corneliussen, U. \ 1998, \aap, 329, 249 
\bibitem[Kroupa (2002)] {kroupa02} Kroupa, P. \ 2002, Science, 295, 82
\bibitem[Lada (1991)] {lada91} Lada, E. A., Bally, J., \& Stark, A.  A. 1991, \apj,  368, 432 
\bibitem[Larson (1985)] {larson85} Larson, R. B. \ 1985, \mnras, 214, L379
\bibitem[Larson (1992)] {larson92} Larson, R. B. \ 1992, \mnras, 256, 641 
\bibitem[Li (2007)] {li07} Li, D., Velusamy, T., Goldsmith, P. F., \& Langer, W. D. 2007, \apj, 655, 351
\bibitem[Li (2004)] {li04} Li, P. S., Norman, M. L., Mac Low, M.-M.,  \& Heitsch, F. 2004, \apj, 605, 800
\bibitem[Li (2003)] {li03} Li, Y., Klessen, R. S., Mac Low, M.-M. 2003, \apj, 592, 975
\bibitem[Massi (2007)] {massi07} Massi, F., De Luca, M., Elia, D. et al. \ 2007, \aap, 466, 1013
\bibitem[Mookerjea (2004)] {mookerjea04} Mookerjea, B., Kramer, C., Nielbock, M., \& Nyman, L.-\AA   \ 2004, \aap, 426, 119 
\bibitem[Motte (1998)] {motte98} Motte, F., Andr\'{e}, P., Neri, R.  \ 1998, \aap, 336, 150 
\bibitem[Motte (2001)] {motte01} Motte, F., Andr\'{e}, P., Ward-Thompson, D., \& Bontemps, S.  \ 2001, \aap, 372, L41
\bibitem[Murray (1996)] {murray96} Murray, S. D., \& Lin, D. N. C. \ 1996, \apj, 467, 728
\bibitem[Munoz (2007)] {munoz07} Mu\~{n}oz, D. J., Mardones, D., Garay, G. et al. \ 2007, \apj, 668, 906
\bibitem[Onishi (1999)] {onishi9} Onishi, T., Mizuno, A., Kawamura, A.,  et al.  \  2002, \apj, 575, 950 
\bibitem[Nakano (1978)] {nakano78} Nakano, T. \& Nakamura, T. \ 1978, \pasj, 30, 671
\bibitem[Padoan (2002)] {padoan02} Padoan, P., \& Nordlund, \AA.  \ 2002, \apj, 576, 870 
\bibitem[Padoan (2007)] {padoan07} Padoan, P., Nordlund, \AA., Kritsuk, A., et al. \ 2007, \apj, 661, 972
\bibitem[Reid (2005)] {reid05} Reid, M. A., \& Wilson, C. D.  \ 2005, \apj, 625, 891
\bibitem[Reid (2006)] {reid06} Reid, M. A., \& Wilson, C. D. \ 2006, \apj, 650, 970  
\bibitem[Salpeter (1955)] {salpeter55} Salpeter, E. E.  \ 1955, \apj, 121, 161 
\bibitem[Scalo (1998)] {scalo98} Scalo, J., The Stellar Initial Mass Function, ed. G. Gilmore \& D. Howell, \ 1998, ASP Conf. Ser., 142, 201  
\bibitem[Shu (2004)] {shu04} Shu, F. H., Li, Z.-Y., \& Allen, A. \ 2004, \apj, 601, 930 
\bibitem[Stanke (2006)] {stanke06} Stanke, T., Smith, M. D., Gredel, R., \& Khanzadyan, T. \ 2006, \aap, 447, 609
\bibitem[Stone (1998)] {stone98} Stone, J. M., Ostriker, E. C., \& Gammie, C. F. \ 1998, \apj, 508, L99
\bibitem[Stutzki (1990)] {stutzki90} Stutzki, J., \& G\"{u}tsen, R. \ 1990, \apj,  356, 513
\bibitem[Testi (1998)] {testi98} Testi, L., \& Sargent, A. I. \ 1998, \apj, 508, L91
\bibitem[Tothill (2002)] {tothill02} Tothill, N. F. H., White, G. J., Matthews, H. E. et al. \ 2002, \apj, 580, 285 
\bibitem[Truelove (1997)] {truelove97} Truelove, J. K., Klein, R. I., McKee, C. F. et al. \ 1997, \apj, 489, L179
\bibitem[Wardthompson (1999)] {wardthompson99} Ward-Thompson, D., Motte, F., Andr\'{e}, P. \ 1999, \mnras, 305, 143 
\bibitem[Wilson (2003)] {wislon03} Wilson, C. D., Scoville, N., Madden, S. C., \& Charmandaris, V. \ 2003, \apj, 599, 104
 
\end{thebibliography}
\end{document}